\RequirePackage{fix-cm}
\RequirePackage{amsmath}
\documentclass[twocolumn,epjc3]{svjour3}
\newcommand{\gerda}       {\textsc{Gerda}}
\newcommand{\gesix}       {{$^{76}$Ge}}
\newcommand{\qbb}         {{$Q_{\beta\beta}$}}
\newcommand{\Th}          {$^{228}$Th}
\newcommand{\onbb}        {{$0\nu\beta\beta$}}
\usepackage{upgreek}     
\newcommand{\thzza}       {{$^{228}$Th}}
\newcommand{\nuc}[2]      {{$^{#2}$\rm #1}}
\newcommand{\Tl}          {$^{208}$Tl}
\newcommand{\Qbb}         {{$\text{Q}_{\beta\beta}$}}

\smartqed  
\RequirePackage{graphicx}

\usepackage[utf8]{inputenc}
\usepackage[T1]{fontenc}
\usepackage{todonotes}
\usepackage{xspace}
\usepackage{multirow}
\usepackage[scientific-notation=true]{siunitx}
\usepackage{physics}
\usepackage{hyperref}
\usepackage{wasysym}

\usepackage{lineno}

\usepackage{float}
\floatstyle{plaintop}
\restylefloat{table}
\usepackage[tableposition=top]{caption}

\graphicspath{{./images/}}

\journalname{Eur. Phys. J. C}
\begin{document}

\title{Calibration of the \gerda\ experiment
}

\author{
The \mbox{\protect{\sc{Gerda}}} collaboration\thanksref{corrauthor}
\and  \\[4mm]
M.~Agostini\thanksref{UCL,TUM} \and
G.~Araujo\thanksref{UZH} \and
A.M.~Bakalyarov\thanksref{KU} \and
M.~Balata\thanksref{ALNGS} \and
I.~Barabanov\thanksref{INRM} \and
L.~Baudis\thanksref{UZH} \and
C.~Bauer\thanksref{HD} \and
E.~Bellotti\thanksref{MIBF,MIBINFN} \and
S.~Belogurov\thanksref{ITEP,INRM,alsoMEPHI} \and
A.~Bettini\thanksref{PDUNI,PDINFN}
% ...
}
\authorrunning{the \textsc{Gerda} collaboration}
\thankstext{corrauthor}{
  \emph{correspondence}  gerda-eb@mpi-hd.mpg.de}
\thankstext{alsoMEPHI}{\emph{also at:} NRNU MEPhI, Moscow, Russia}
\thankstext{alsoLev}{\emph{also at:} Moscow Inst. of Physics and Technology,
  Russia}
\thankstext{nowIKZ}{\emph{present address:} Leibniz-Institut f{\"u}r
  Kristallz{\"u}chtung, Berlin, Germany}
\thankstext{alsoDubna}{\emph{also at:} Dubna State University, Dubna, Russia}
\thankstext{nowMacolino}{\emph{present address:} LAL, CNRS/IN2P3,
  Universit{\'e} Paris-Saclay, Orsay, France}
\institute{
INFN Laboratori Nazionali del Gran Sasso and Gran Sasso Science Institute, Assergi, Italy\label{ALNGS}
% ...
\and
Max-Planck-Institut f{\"u}r Kernphysik, Heidelberg, Germany\label{HD} \and
Department of Physics and Astronomy, University College London, London, UK\label{UCL} \and
Dipartimento di Fisica, Universit{\`a} Milano Bicocca, Milan, Italy\label{MIBF} \and
INFN Milano Bicocca, Milan, Italy\label{MIBINFN}
% ...
\and
Institute for Nuclear Research of the Russian Academy of Sciences, Moscow, Russia\label{INRM} \and
Institute for Theoretical and Experimental Physics, NRC ``Kurchatov Institute'', Moscow, Russia\label{ITEP} \and
National Research Centre ``Kurchatov Institute'', Moscow, Russia\label{KU}
% ...
\and
Physik Department, Technische  Universit{\"a}t M{\"u}nchen, Germany\label{TUM} \and
Dipartimento di Fisica e Astronomia, Universit{\`a} degli Studi di
Padova, Padua, Italy\label{PDUNI} \and
INFN  Padova, Padua, Italy\label{PDINFN}
% ...
\and
Physik-Institut, Universit{\"a}t Z{\"u}rich, Z{u}rich, Switzerland\label{UZH}
}

\maketitle

\begin{abstract}
  The GERmanium Detector Array (\gerda) collaboration searched for neutrinoless double-\(\beta\) decay in \gesix\ with an array of about 40 high-purity isotopically-enriched germanium detectors.
  The experimental signature of the decay is a monoenergetic signal at \qbb$=2039.061(7)$\,keV in the measured summed energy spectrum of the two emitted electrons.
  Both the energy reconstruction and resolution of the germanium detectors are crucial to separate a potential signal from various backgrounds, such as neutrino-accompanied double-\(\beta\) decays allowed by the Standard Model.
  The energy resolution and stability were determined and monitored as a function of time using data from regular \Th\ calibrations.
  In this work, we describe the calibration process and associated data analysis of the full \gerda\ dataset, tailored to preserve the excellent resolution of the individual germanium detectors when combining data over several years.
\end{abstract}

\section{Introduction}
\label{sec:intro}
Neutrinoless double-\(\beta\) (\onbb) decay is a hypothetical, second-order weak interaction process in which a nucleus changes its charge number by two units with the emission of two electrons but without accompanying anti-neutrinos.
This lepton-number violating process is only permitted if neutrinos are massive Majorana fermions, i.e. if there is a Majorana mass term in the Lagrangian of the underlying theory.
Such a term appears in many extensions of the Standard Model of particle physics and could explain why neutrino masses are much smaller than those of all other fermions~\cite{Mohapatra:2006gs}.
%
            The GERmanium Detector Array (\gerda) collaboration searched for the \onbb\ decay of the isotope \gesix\ with a $Q$-value of \linebreak${Q_{\beta\beta}=2039.061(7)}$\,keV~\cite{Mount:2010zz} by operating high-purity germanium (HPGe) detectors isotopically enriched to $>$86\% in \gesix, making them also the potential source of \onbb\ decay.

We used three types of enriched germanium detectors: 30 broad energy germanium (BEGe) detectors, 7 coaxial detectors, and 5 newer inverted coaxial (IC) detectors.
The BEGe detectors are smaller (average 0.7\,kg) but offer superior energy resolution and pulse shape discrimination (PSD) properties compared to the coaxial detectors~\cite{Agostini:2019mwn}, while the IC detectors provide energy resolution and PSD properties similar to the BEGe detectors~\cite{Domula:2017mei} but with a larger mass (average 1.9\,kg) comparable to that of the coaxial detectors (average 2.3\,kg), allowing for the easier design of larger germanium arrays.

The array of germanium detectors was immersed in a cryostat filled with 64\,m$^3$ of liquid argon (LAr). The top of the cryostat and the surrounding water tank houses a clean room containing a glove box and lock system for deploying the HPGe detectors and calibration sources. The entire setup was located underground at the Laboratori Nazionali del Gran Sasso (LNGS) of INFN, Italy, and is described in detail in~\cite{Agostini:2017hit}.

The first phase of the experiment was operated with 18\,kg of coaxial detectors (inherited from the Heidelberg-Moscow~\cite{Gunther:1997ai} and \textsc{Igex}~\cite{Morales:1998hu} collaborations) between November 2011 and September 2013. Phase~II started in December 2015, after 20\,kg of BEGe detectors produced for the \gerda\ experiment were added and the liquid argon volume around the detector array was instrumented with photosensors as a veto against radioactivity~\cite{Agostini:2017hit}. During an upgrade in mid-2018, referred to as the Phase~II upgrade, IC detectors with a total mass of 9.6\,kg were added, and the LAr instrumentation was upgraded. Phase~II ended in November 2019.
While the calibration procedure of Phase~I data has been discussed in \cite{Ackermann:2012xja,Agostini:2013mzu}, the focus of this paper is the calibration of the Phase~II data.

In all recent \onbb\ decay experiments, the signature of the rare nuclear transition is a monoenergetic peak in the measured energy spectrum of the two electrons at \qbb.
Consequently, a crucial parameter to distinguish a signal from the background is the energy estimator.
The better the energy resolution of the detectors, the narrower the signal energy region effectively becomes, and an excess over the continuous background can be more clearly identified.
One strength of HPGe detectors is their unparalleled energy resolution (typically $\sigma$/E$\sim$0.1\% at \qbb).
It permits the almost complete rejection of background events from regular two-neutrino-accompanied double-\(\beta\) decays~\cite{Abramov:2019hhh}, an otherwise irreducible background in \onbb\ decay searches~\cite{Elliott:2002xe,Maneschg:2017mzu}.

Given the central role of the energy observable, adequate measures must be taken to accurately determine the energy scale and resolution, monitor their stability over the full data acquisition period, and determine the relevant uncertainties entering the statistical analysis for the \onbb\ decay search.
In Sect.~\ref{sec:calprocess} we detail the calibration procedure, while in Sect.~\ref{sec:calibProcessing} we discuss the analysis of the calibration data and the energy scale determination, including the procedures employed to monitor and maintain the stability of the HPGe detectors over time.
In Sect.~\ref{sec:combinedanalysis} we describe the determination of the energy resolution for the \onbb\ decay analysis, and in Sect.~\ref{sec:sys_res} we provide an evaluation of the associated uncertainties.
In Sect.~\ref{sec:sysscale}, we discuss the determination of the residual energy bias and its uncertainty.
In Sect.~\ref{sec:compphy}, we compare the results from calibration data with those in the physics data (data used for the \onbb\ decay search) for the resolutions of the lines from decays of $^{40}$K and $^{42}$K.
We close in Sect.~\ref{sec:conclusions} with a summary and a discussion of our main results.

\section{Energy calibration process}
\label{sec:calprocess}

To perform the calibrations, we regularly exposed the HPGe detectors to three custom-made low-neutron emission \Th\ calibration sources~\cite{Baudis:2015sba}, each with an activity of about 10\,kBq.
These sources were stored within shielding above the lock system, at a vertical distance of at least 8\,m to the HPGe detector array, during physics data acquisition.
Since \Th\ has a half-life of 1.9\,yr, the sources were replaced during Phase~II to ensure a sufficient level of activity.

During calibration runs of the HPGe detectors, the \Th\ sources were lowered into the LAr cryostat to reach the level of the detector array by three source insertion systems~\cite{tarkathesis,baudis2013monte}.
Each of these deploy a single source, placed on tantalum absorbers (h = 60\,mm, $\diameter$ = 32\,mm)
During calibration, each source was placed at three different heights to expose the detector array more homogeneously, and data were acquired at each location for up to 30\,min. With this exposure, typically around $(1-3)\times10^3$ events are observed in the prominent $^{208}$Tl $\gamma$ line at 2614.5\,keV in a BEGe detector, and $(0.6-1)\times10^4$ events in a coaxial or IC detector. 
Calibration data were acquired every 7-10 days with a total of 142 calibration runs used for the analysis of the Phase~II data.

The triggering energy threshold for this acquisition during calibration corresponds to $\sim$400\,keV.
This threshold was set to include the strong $\gamma$ line of the \Th\ spectrum at 583.2\,keV while keeping the event rate at a manageable level for the data acquisition system.
The detector signals were read out with charge sensitive amplifiers, and digitized by a 100\,MHz 14-bit flash analog-to-digital converter (FADC).
As for physics data acquisition, for each trigger a 160\,$\upmu$s long waveform is recorded at a sampling rate of 25\,MHz, centered around the trigger time and covering an energy range up to $\sim$6\,MeV.
During calibration, every 2\,s a test pulse was injected into the amplifier electronics of all germanium detectors to monitor the stability of their gains.
Between successive calibration runs, i.e. during physics data acquisition, test pulsers were injected every 20\,s for the same purpose.

Data from the FADCs are transformed into the MGDO (ROOT-based) format~\cite{Brun:1997pa,Agostini:2011nf} and processed to analyze properties of the recorded waveforms using the {\sc{Gelatio}} software~\cite{Agostini:2011xe}, as is the case for physics data. The energy is estimated from the amplitude of the waveform after applying a digital filter which reduces the impact of noise and thus improves the resolution. As a fast first estimate for monitoring and cross-check purposes, a pseudo-Gaussian filter is applied to obtain an energy estimator as part of the online analysis~\cite{gatti1986processing}.
An improved energy resolution for the final data analysis is achieved with a Zero Area Cusp (ZAC) filter~\cite{Agostini:2015pta}, which removes the effect of low-frequency noise.
This filter is optimized offline for each calibration run and HPGe detector to minimise the resolution of the highest energy $\gamma$ line in the \Th\ spectrum~\cite{valerio:thesis}.

A set of heuristic event selection criteria is applied to ensure that events recorded during calibration are of a physical origin, and to reduce pile-up events. The underestimation of energy for these events cause low-energy tails in the spectra of $\gamma$ lines and can bias the estimated energy resolution. These selection criteria are based on the properties of the waveform, such as the baseline stability and slope, trigger time, number of triggered events, and rise time of the pulse. The probability of rejecting physical interactions, estimated with events from the regularly injected test pulses, is below 0.1\%~\cite{lazzarothesis}.

\section{Analysis of energy spectra}
\label{sec:calibProcessing}

\begin{figure*}[tb]
\centering
\includegraphics[width=\textwidth]{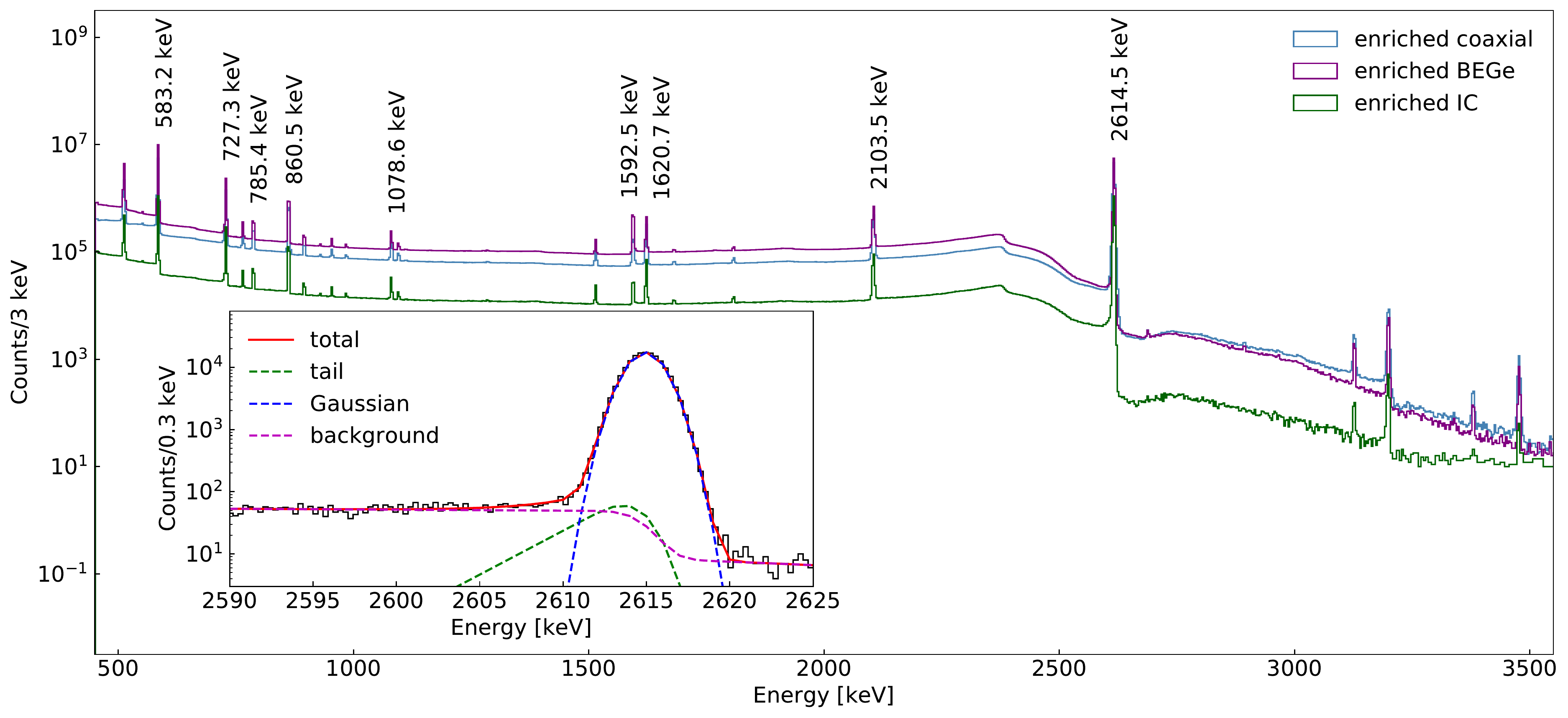}
\caption{Combined energy spectrum for \thzza\ calibration data for all enriched detectors of BEGe, coaxial, and IC type during Phase~II after rebinning to 3\,keV. The inset shows the fit to the 2.6\,MeV line in the spectrum of the detector GD91A before the Phase~II upgrade with 0.3\,keV binning, with the components of the fit drawn separately (linear and step backgrounds are combined). The energies of the nine peaks that typically contribute to the formation of calibration curves are labelled.}
\label{fig:supercalib}
\end{figure*}

Nuclei of the \thzza\ isotope decay in a chain via $\alpha$ and $\beta$ decays to the stable \nuc{Pb}{208}\ with the emission of multiple monoenergetic $\gamma$ rays. These result in sharp peaks in the recorded energy spectra, as shown in Fig.~\ref{fig:supercalib} in the combined spectra of each detector type. The pattern of observed peaks is used to identify the $\gamma$ lines and thereby determine their energy and resolution.
The \texttt{TSpectrum} class of ROOT is used to find peak positions in the uncalibrated spectrum, such that all peaks with amplitudes exceeding 1/20 of the amplitude of the most prominent peak are found.
This threshold was chosen to avoid the detection of spurious peaks.
The peak with the highest energy is identified as the full energy peak (FEP) of the $\gamma$ ray from the decay of \Tl, a daughter of \Th, at $E_{\mathrm{FEP}}=2614.5$\,keV. A preliminary calibration for the energy estimator $T$ is applied assuming a linear energy scale without offset:

\begin{align}
E_0(T)=\frac{E_{\mathrm{FEP}}}{T_{\mathrm{FEP}}}\cdot T\,.
\end{align}
A candidate peak is confirmed if its preliminary estimated energy is consistent within 6\,keV with the energy of a known line in the \Th\ spectrum.
The 6\,keV value permits the accurate identification of peaks while allowing for some non-linearity of the energy scale.
The known peaks correspond to $\gamma$ rays from isotopes in the \Th\ decay chain with energies above 500\,keV and branching ratios above 0.3\%,
including the detector specific single escape peak (SEP) at 2103.5\,keV and double escape peak (DEP) at 1592.5\,keV resulting from the 2.6\,MeV $\gamma$ ray of \Tl decays.
In the context of this paper, without ambiguity, FEP, SEP, and DEP always refer to those of \Tl.
The double peak due to the 511.0\,keV annihilation line and 510.7\,keV $\gamma$ line from $^{208}$Tl is excluded from the analysis, in particular since the resolution of the annihilation peak is broadened due to the Doppler effect~\cite{coldwell2016experimental}.

\subsection{Peak fitting and calibration curves}
\label{subsec:peak_params}

To determine the position $\mu$ and energy resolution in terms of $\textrm{the full width at half maximum (FWHM)}=2.35\cdot\sigma$ of the identified peaks, fits are performed locally in an energy window of $10-20$\,keV around the peak position obtained from the preliminary calibration. These are configured manually and separately for each peak to avoid interference from neighbouring peaks.

Minimally, a Gaussian $g(E)$ is used to model the peak, and a linear function $f_\mathrm{lin}(E)$ is used to model the background:

\begin{align}
  g(E) &= \frac{n}{\sqrt{2\pi}\sigma}\exp\qty[-\frac{(E-\mu)^2}{2\sigma^2}], \label{eq:gauss} \\
  f_{\mathrm{lin}}(E) &= a + b \cdot E, \label{eq:bglin}
\end{align}
where $n$, $\mu$ and $\sigma$ are the intensity, position, and width of the peak, and \(a\) and \(b\) give the intercept and slope of the linear function, respectively.

\sloppy For high statistics peaks (583.2\,keV, 727.3\,keV, 763.5\,keV, 860.5\,keV, and 2614.5\,keV), the SEP, and the DEP, a step function is used to model the flat backgrounds occurring only above or below the peak from multiple Compton scatters:

\begin{align}
  f_{\mathrm{step}}(E)  = \frac{d}{2} \mathrm{erfc} \left( \frac{E-\mu}{\sqrt{2}\sigma} \right),
  \label{eq:bgstep}
\end{align}
where \(d\) is the height of the step function, and erfc denotes the complementary error function.

For the high statistics peaks as defined above, a low-energy tail is additionally used to model the effects of incomplete charge collection and the residual presence of pile-up events:

\begin{align}
  h(E) = \frac{c}{2\beta}\exp\left(\frac{E-\mu}{\beta}+\frac{\sigma^2}{2\beta^2}\right)\mathrm{erfc}\left(\frac{E-\mu}{\sqrt{2}\sigma}+\frac{\sigma}{\sqrt{2}\beta}\right),
  \label{eq:tail}
\end{align}
where \(\beta\) and \(c\) are the height and slope of the tail, respectively.
An example of the FEP peak fit is shown in the inset of Fig.~\ref{fig:supercalib}.

Peaks are excluded after the fit if any of the following heuristic rules are fulfilled: (i) the estimated FWHM is above 11\,keV or below 1.5\, keV; (ii) the peak maximum is lower than 2.5 times the linear component of the background or lower than 10 counts; (iii) the fitting error on the FWHM is larger than the FWHM itself.
These rules are purely heuristic and designed to remove peaks that cannot be fitted well, mainly due to low statistics.

Typically around 5-8 peaks per detector survive all selection criteria. The FEP is always identified, since the peak identification algorithm requires it. In >80\% of cases the lines at 583.2\,keV, 860.5\,keV, 1592.5\,keV (DEP) and 2103.5\,keV (SEP), and in (15-60)\% the lines at 727.3\,keV, 785.4\,keV, 1078.6\,keV and 1620.7\,keV are found. All other $\gamma$ lines are seen in $<$3\% of the spectra from individual detectors of a single calibration run, due to insufficient statistics.

From the obtained peak positions, we determine the calibration curve which is a function to convert the uncalibrated energy estimator $T$ into a physical energy in keV.
We plot the peak positions in terms of the uncalibrated energy estimator $T$ of identified peaks against their physical energies \(E\) according to literature values~\cite{litvals}, and then fit with a linear function
\begin{align}
E(T)=p_0+p_1\cdot T.
\end{align}
In the rare case where only the FEP is found and successfully fitted, the resulting calibration curve has an intercept of zero.
Such cases only occured during periods of instability for a detector which were excluded from data analysis (see Sect.~\ref{sec:monitoring}).
For most detectors, the linear calibration curve describes the peak positions within a few tenths of a keV, as shown in Sect.~\ref{sec:sysscale}.
Discussion of a quadratic correction to the energy scale introduced for five detectors after the Phase~II upgrade can be found in Sect.~\ref{ssec:quadcorr}.

Typically, a calibration curve is used to calibrate the physics data following a calibration run.
However, if a calibration is taken after changes in the experimental setup, or instabilities in the detector array, the calibration curves may be applied retrospectively to additionally calibrate the period between the changes or instabilities and the calibration run.
The unstable period itself would not be used for physics analysis.

\subsection{Quadratic correction}
\label{ssec:quadcorr}
After the Phase~II upgrade, several detectors (the new IC detectors and one coaxial detector) exhibited larger residuals in their calibration curves compared to other detectors, up to 2.5\,keV at 1.5\,MeV.
Whether these effects are directly related to change in cable routing during the Phase~II upgrade, is unknown.
These effects could be largely accounted for by the incorporation of a quadratic correction to the calibration curves,

\begin{align}
E(T)=m_0+m_1\cdot E_0(T) + m_2\cdot E_0^2(T),
\end{align}
where $E_0$ is the energy estimator after the application of the linear calibration curve as described in Sect.~\ref{subsec:peak_params}.

\begin{figure}
  \centering
  \includegraphics[width=.5\textwidth]{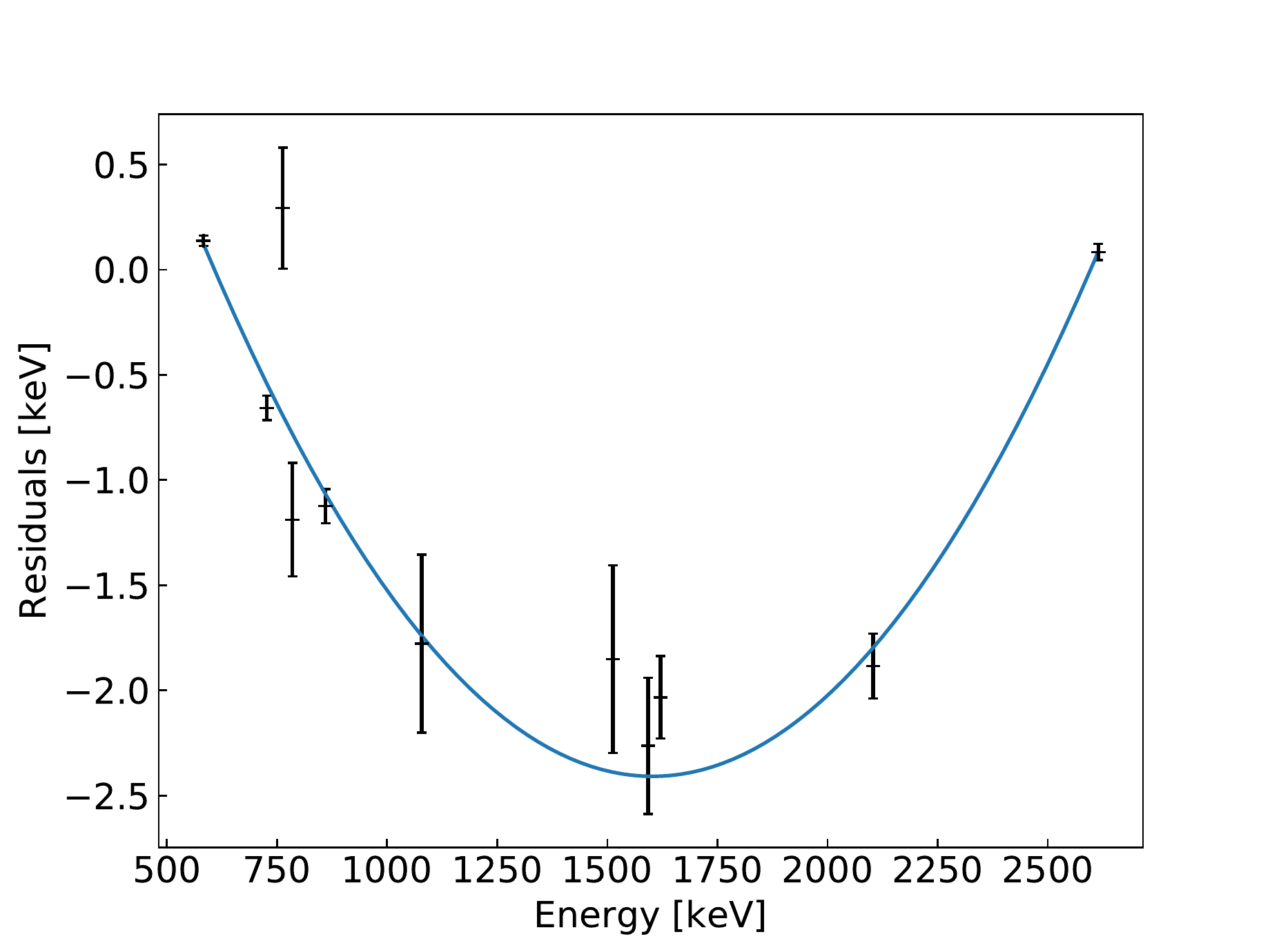}
\caption{Fitting the residuals of the calibration curve with a quadratic function, as shown for detector ANG2 for the calibration on 15th October 2018. }
  \label{fig:fit_quad}
\end{figure}

The parameters $m_0$, $m_1$ and $m_2$ were determined by fitting the residuals of each detector's calibration curves per calibration run. One such example is shown in Fig.~\ref{fig:fit_quad}. These parameters were observed to be stable over time, with the exception of a single jump for three IC detectors following hardware activities in the clean room.
Therefore a single correction for each detector's stable period was applied, using the average parameters of all calibration runs in that period.
After the quadratic correction, the remaining residuals were within a few tenths of a keV.

\subsection{Detector performance and stability}
\label{sec:monitoring}
To consistently combine data over an extended period of time while preserving the excellent energy resolution of the HPGe detectors,
it is vital to monitor the stability of the energy scale between calibrations and exclude periods with significant shifts and fluctuations which would contribute to the width of the peaks.
As previously mentioned, test pulses are regularly injected into the readout electronics to monitor the stability of the data acquisition system. Their signal magnitude corresponds to an energy of about 3\,MeV. Periods with significant jumps or drifts (>1\,keV) in the amplitude of the test pulses are excluded from data analysis and a calibration is performed once the detector stabilizes.
The corresponding loss of exposure is at the few-percent level.
The origin of these shifts is largely unknown, but may be caused by temperature changes in the electronics.

Additionally, we monitor the stability of the FEP position in the calibration spectrum. 
If the position of this line changes by more than 1\,keV between successive calibrations without an identifiable reason (maintenance, longer breaks, specific incident), the data of the respective detector are discarded from the analysis for that period of time.
The corresponding exposure loss is at the few-percent level.
Smaller or temporary drifts may still affect the obtainable energy resolution and are discussed as a systematic uncertainty in Sect.~\ref{sec:pulserstab}.

\begin{figure}
  \centering
  \includegraphics[width = 0.5\textwidth]{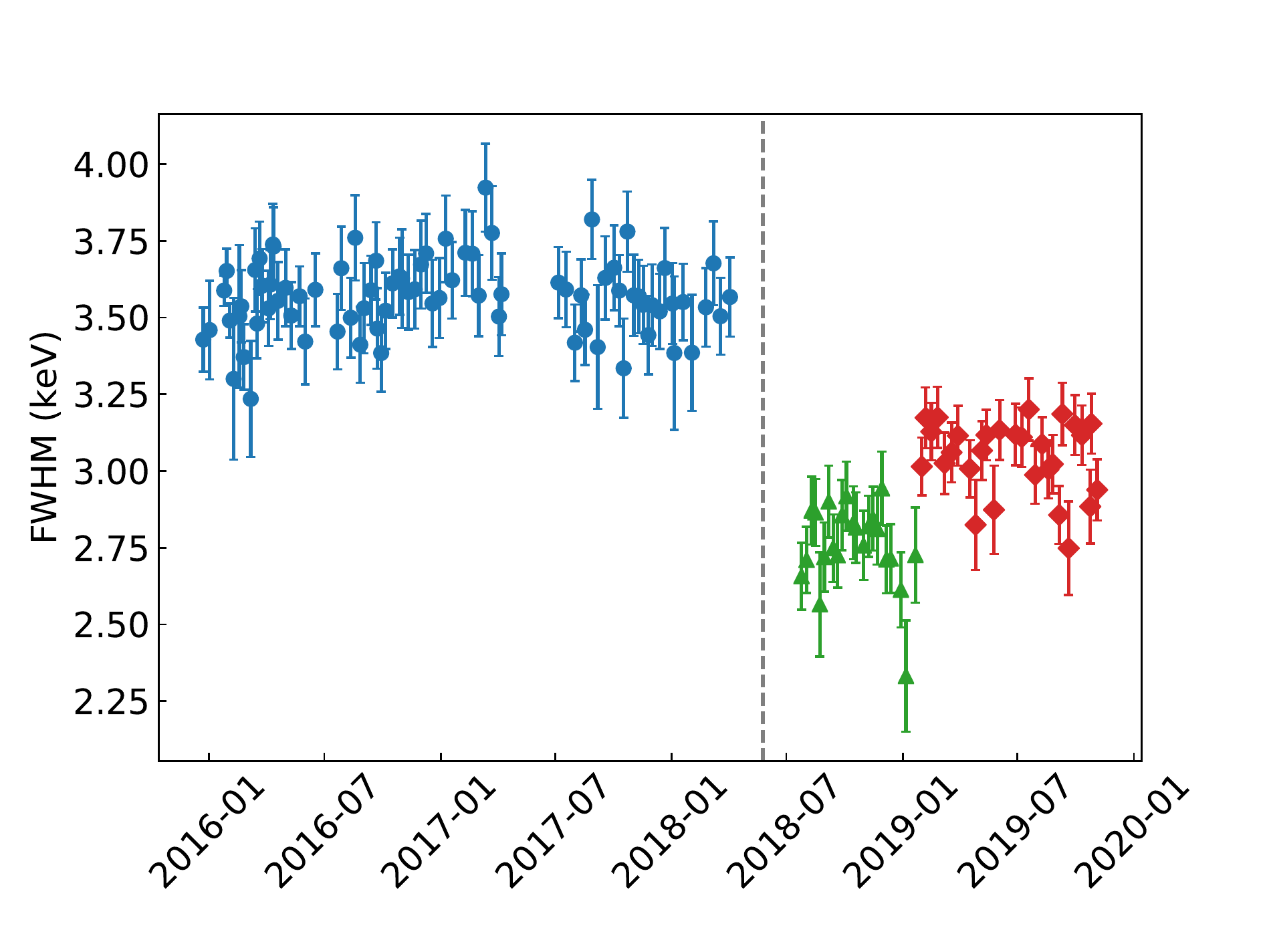}
  \caption{FWHM of the FEP as a function of time for detector GD76B, one of the BEGe detectors. Each data point comes from one calibration run. The full data acquisition period is divided into three partitions, shown in solid circle (blue), triangle (green), and diamond (red), respectively. The partition shown in triangles is due to the Phase II upgrade and coincident improvement in resolution. The partition shown in diamonds is due to the jump in resolution in January 2019 when a hardware change took place.}
  \label{fig:fwhm_partition}
\end{figure}

Due to hardware changes, the detectors may experience changes in their energy resolution and energy scale over longer periods of time.
To more accurately reflect the properties of a detector at a certain time, for the final \gerda\ analysis \cite{Agostini:2020xta} we divide the full data acquisition period for each detector into stable sub-periods called partitions. The stability is judged based on two parameters: the FWHM at the FEP and the residual at SEP.
The former reflects the changes in the detector resolution, while the latter catches the changes in the energy bias at the energy peak closest to $Q_{\beta \beta}$ (see Sect.~\ref{sec:sysscale} for more discussions on the bias).
After the Phase~II upgrade and cable rerouting, the resolutions improved for most of the detectors.
Therefore, for simplicity, we start a new partition for all detectors after the upgrade.
There are one to four partitions for each detector. The majority of the detectors have only two partitions, split at the time of the Phase~II upgrade.
An example of the partitions is shown in Fig.~\ref{fig:fwhm_partition}.

\section{Energy resolutions from the  combined calibration spectra}
\label{sec:combinedanalysis}
Depending on the specific physics analysis, we calculated the energy resolution either by partition, described in the previous section, or by detector type. For the \onbb\ decay search reported in~\cite{Agostini:2019hzm}, where the data before the Phase~II upgrade were analyzed, an effective resolution for each detector type was employed. For the final \onbb\ decay search of \gerda\ reported in~\cite{Agostini:2020xta}, where all \gerda\ data were analyzed, we calculated a resolution for each partition, a much more fine-grained approach. At the expense of increased complexity, the partition approach improves the physics result by capturing the variations among the detectors as well as the variation over time.

Since both methods are applicable for \gerda\, and any other experiment with a modular detector setup, here,
we discuss both approaches.
While the detector type approach was replaced in favor of partitioning for the final \onbb\ decay search, the former gives an overview of the overall detector performance.
For this reason all the illustrations and calibration parameters are provided by detector type.
For simplicity, we refer to a collection of detectors of the same type as a dataset. 

\subsection{By partition}
\label{ssec:respartition}

To obtain the $\gamma$ line resolutions for each detector partition, we first produce combined calibration spectra.
The energy spectra obtained from each calibration run within one partition are first normalised to account for differing statistics, and then weighted according to the time span for which the corresponding calibration curves were used to calibrate physics data.
The resulting $\gamma$ peaks in a combined spectrum will be representative of the average performance of that detector in that partition. 

The peak identification and fit procedure described in Sect.~\ref{sec:calibProcessing} is then applied to each combined calibration spectrum.

The SEP is broadened due to the known Doppler effect and is thus excluded~\cite{coldwell2016experimental}. We also observe broadening in the DEP. This is hypothesised to originate due to events occurring more frequently in the outer regions of the detectors and thus being more susceptible to incomplete charge collection~\cite{hull1996charge}. This line is therefore excluded as well.

The dependence of the $\gamma$ line resolutions on the calibrated energy $E$ is then fitted with the function~\cite{Agostini:2015pta}

\begin{align}
  \sigma(E) = \sqrt{a + bE},
\label{eq:rescurve}
\end{align}
where $a$ and $b$ are fit parameters.
The former accounts for the contributions from electronic noise, while the latter accounts for statistical fluctuations in the number of charge carriers.
The resolution at \Qbb\ is then given by using \(E = Q_{\beta\beta}\) in Eq.~\ref{eq:rescurve}.

\begin{figure}[htb]
  \centering\includegraphics[width=0.5\textwidth]{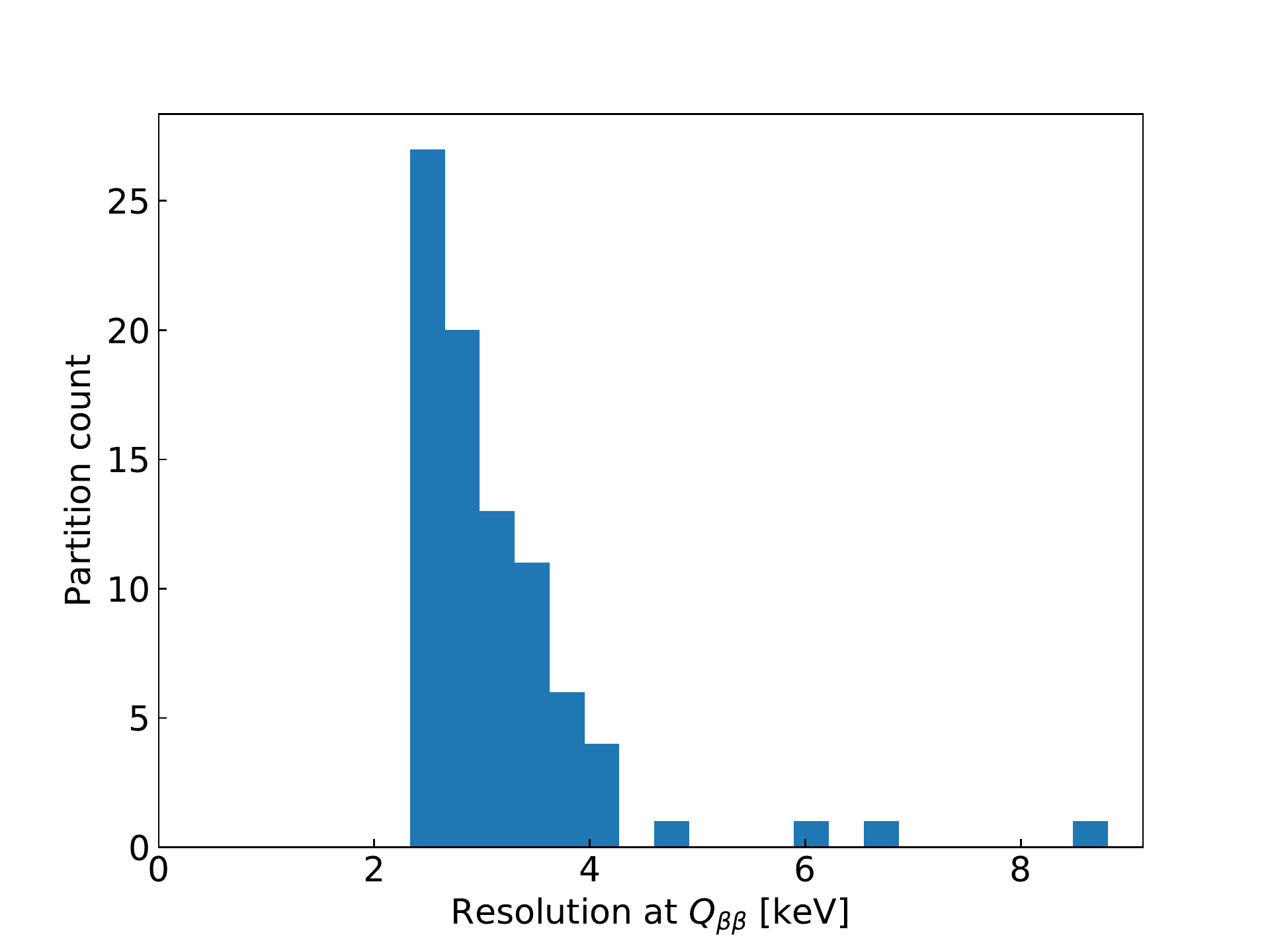}
\caption{Distribution of FWHM resolution at \Qbb\ per detector partition. The detector partitions with resolutions >\,6\,keV are due to two coaxial detectors whose resolutions degraded after the Phase II upgrade. }
\label{fig:resdist}
\end{figure}

The resultant FWHM resolutions at \Qbb\ of the partitions vary between 2.3\,keV and 8.8\,keV, as shown in Fig.~\ref{fig:resdist}.
Values for each partition can be found in \cite{chloethesis}.
Systematic errors are calculated via a dedicated study as explained in Sect.~\ref{sec:sys_res}.

\subsection{By detector type}
\label{ssec:restype}
The appropriate method for calculating effective resolutions by detector type depends on the specific application.

\subsubsection{Background modeling}
\label{sssec:bkgeffres}

For background modeling, energy dependent resolutions are required, i.e. resolution curves.
To calculate these for datasets, the procedure is similar to that for the partitions, though weighting is now required to combine the resolutions from different detector partitions.
When data from multiple partitions are combined by adding their energy spectra, Gaussian peaks in the individual spectra combine to become a Gaussian mixture, namely the sum of multiple Gaussian distributions with different centroids and resolutions.
The resolution of individual partitions in a dataset is stable within a factor of 1.7  for BEGe and IC detectors. 
For coaxial detectors there is a slightly higher fluctuation, but still within a factor of three.
The variation in position of the centroid is much smaller than the energy resolution, typically around 0.2\,keV.
Therefore the shape of a peak in the combined energy spectra remains approximately Gaussian and can be characterized by an effective resolution, computed from the resolution of individual partitions.

The variance of a Gaussian mixture is given by:
\begin{align}
	\mathrm{\sigma}^2 = \sum_i w_i\left(\sigma_i^2 + \mu_i^2\right) - \left(\sum_i w_j\cdot\mu_i \right)^2,
\label{eq:gaussmix}
\end{align}
where the sum goes over Gaussians with means and standard deviations $\mu_i$ and $\sigma_i$, with weights $w_i$,
representing the relative contribution to expected peak counts of individual Gaussians~\cite{gaussianmixturethesis}.

For a dataset comprised of individual partitions, these parameters stand for the individual partitions' resolution $\sigma_i$, and
peak position $\mu_i$, which can be different due to independent systematic effects on the energy scale.
The weights are the expected relative event count contribution of individual partitions.
Since peak counts are proportional to exposure $\mathcal{E}_i=m_i\cdot t_i$, with individual detector's active mass $m_i$ and live time $t_i$, the relative exposure contribution is:

\begin{align}
w_{i} = \frac{\mathcal{E}_i}{\mathcal{E}},
\label{eq:weight}
\end{align}
where $\mathcal{E}=\sum_j \mathcal{E}_j$ is the total exposure of the dataset.

Since the biases in the energy scale are small, we can neglect the differences in the peak positions.
Eq.~\ref{eq:gaussmix} therefore simplifies to:

\begin{align}\label{eq:simpgaussmix}
	\sigma = \sqrt{\frac{1}{\mathcal{E}}\sum_i \mathcal{E}_i \,\sigma_i^2},
\end{align}
with total error $\delta_{\sigma}$ from the statistical fitting errors of individual partition resolutions $\delta_{\sigma_i}$:

\begin{align}\label{eq:gaussmixerr}
	\delta_{\sigma} = \sqrt{\frac{1}{\mathcal{E}^2\sigma^2}\sum_i (\mathcal{E}_i \sigma_i \delta_{\sigma_i})^2},
\end{align}
with negligible uncertainty in the weights.

\begin{figure*}[ht!]
  \centering\includegraphics[width = \textwidth]{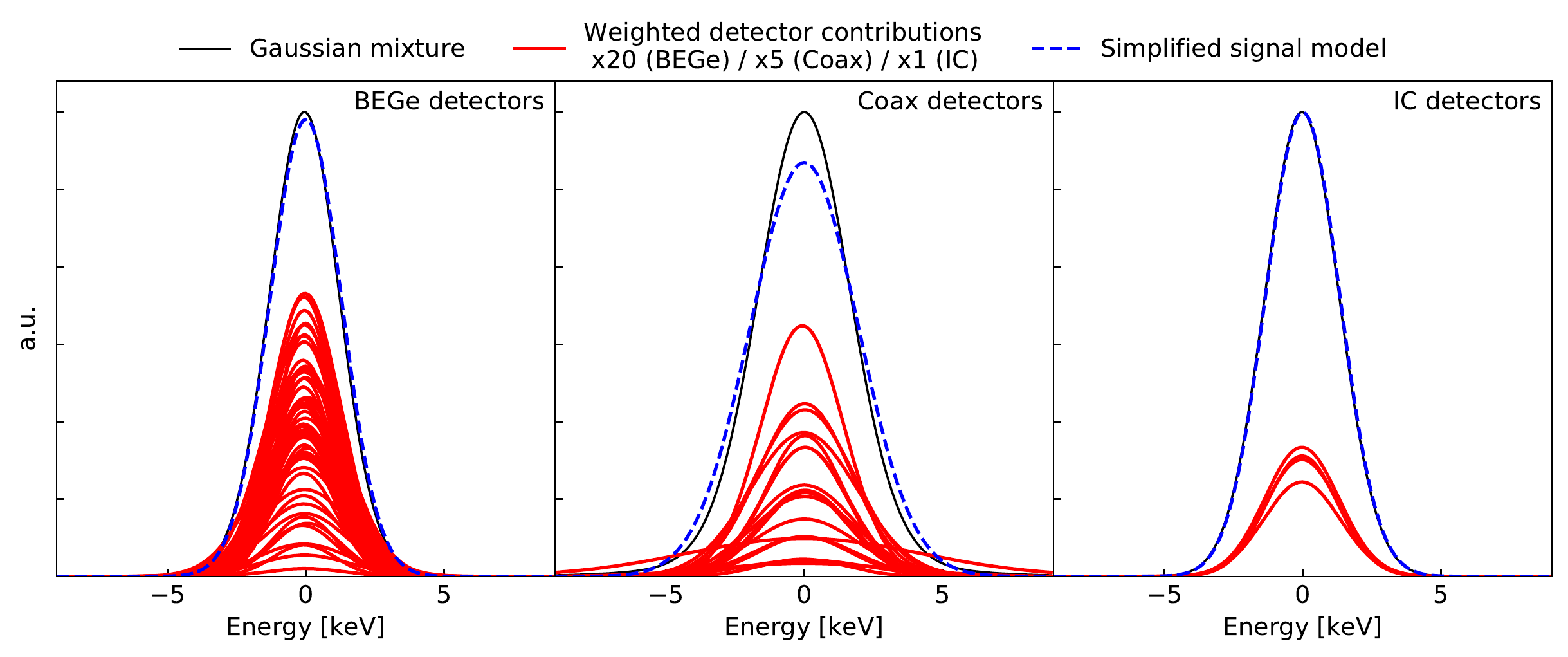}
  \caption{Comparison of simplified Gaussian signal model (dashed blue) to the more detailed Gaussian mixture signal model (solid black) of the FEP, for datasets formed of the partitions of BEGe (left), coaxial (middle) and IC (right) detectors.
    Red lines show Gaussian shaped peaks for individual partitions, which have been rescaled by a factor of 20/5/1 for the BEGe/Coax/IC detectors for visibility.
 }
 \label{fig:ds-signalmodel}
\end{figure*}

For instance, the simplified model of the FEP is a Gaussian with a mean of 2614.5\,keV and a width fixed to the effective resolution (see Eq.~\ref{eq:simpgaussmix}) of the dataset.
On the other hand, a Gaussian mixture model would consist of the sum of a Gaussian for each partition, each with its own resolution and centroid.
Fig.~\ref{fig:ds-signalmodel} shows the Gaussian mixture and simplified signal models for the IC and coaxial datasets.
For the IC and BEGe datasets, the Gaussian mixture model is very close to a Gaussian shape, as the centroid differences are small and the partitions in each dataset have similar resolutions.
The resolutions among the coaxial detectors are more varied and thus using a Gaussian signal model may be less appropriate.

To calculate the effective resolution curves for each dataset, first the $\gamma$ line resolutions are obtained for each of the partitions as in Sect.~\ref{ssec:respartition}.
For all $\gamma$ lines whose resolution was reliably determined for all partitions in that dataset, an effective resolution of the dataset at that energy is calculated using Eq.~\ref{eq:simpgaussmix}.
All other lines which were missing in at least one detector partition are excluded.

Once the effective resolutions for each energy and dataset have been determined by weighting partition resolutions with Eq.~\ref{eq:simpgaussmix}, their energy dependence is fitted with Eq.~\ref{eq:rescurve}.

\begin{figure}[htb]
  \centering\includegraphics[width = 0.5\textwidth]{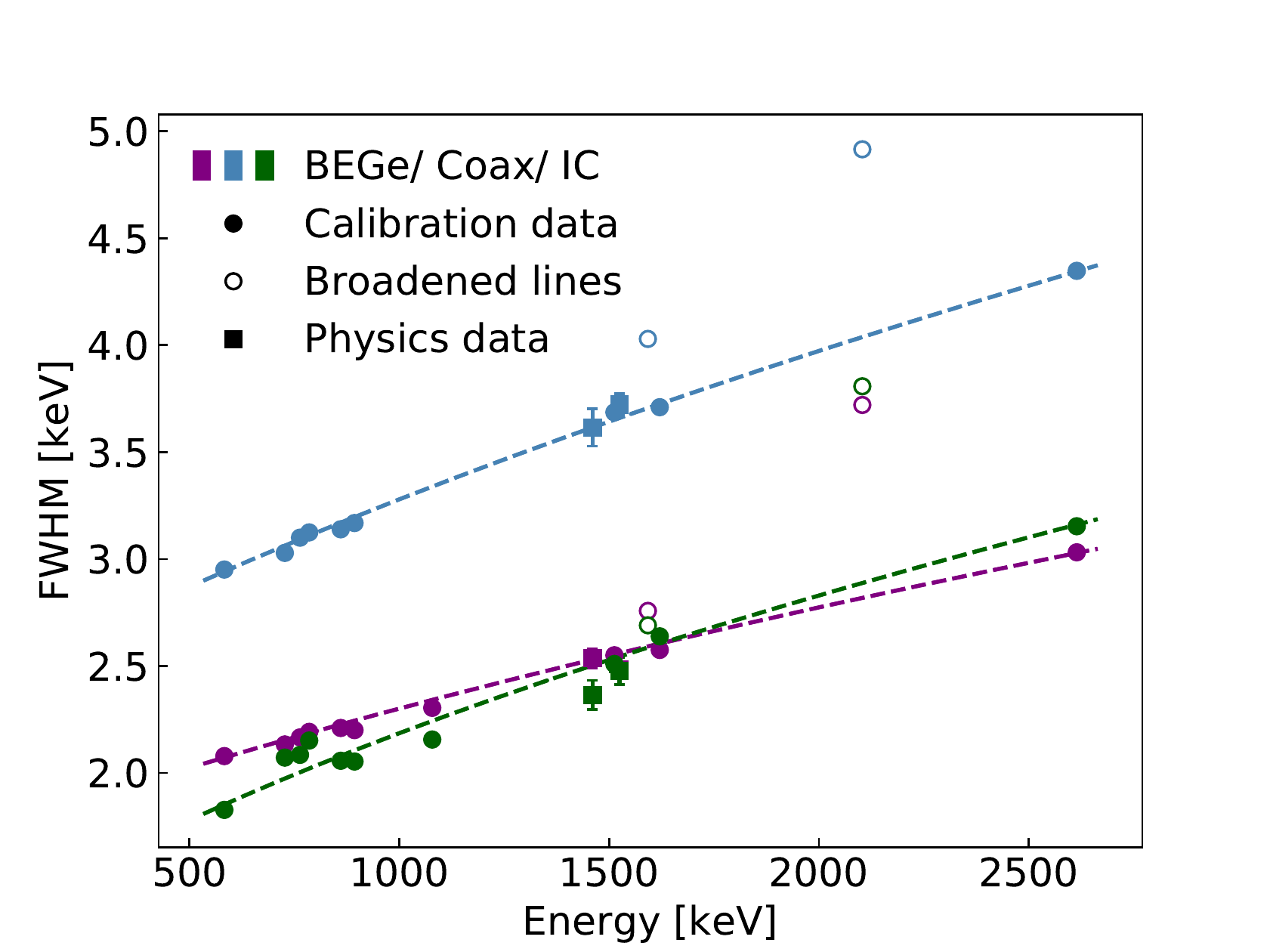}
\caption{Effective resolution curves for BEGe (purple), coaxial (blue) and IC (green) datasets. Open points indicate broadened lines not used to form the resolution curves, namely the double- and single-escape peaks of the 2.6\,MeV line due to \Tl\ decay.
  Square markers indicate the exposure-weighted resolutions of the lines in physics data due to $^{40}$K (1460.8\,keV) and $^{42}$K (1524.7\,keV) decays.
}
\label{fig:uncorrsupercal}
\end{figure}

The obtained effective resolutions and functions of the three detector types are shown in Fig.~\ref{fig:uncorrsupercal} and Tab.~\ref{tab:supercalfit}.
The statistical errors are obtained from the fit.

\begin{table}[h]
\begin{tabular}{c | c c c}
\multirow{2}{*}{Detector type} & $a$  & $b$   \\
 & [keV$^2$] & [$10^{-4}$ keV] \\
\hline
BEGe & 0.551(1) & 4.294(9)  \\
coaxial & 0.985(2) & 10.73(2)  \\
IC & 0.280(2) & 5.83(2)
\end{tabular}
\centering\caption{Parameters of resolution curves (Eq.~\ref{eq:rescurve}) obtained for datasets of each detector type.}
\label{tab:supercalfit}
\end{table}

\subsubsection{\onbb\ decay search}
\label{sssec:decayres}

As mentioned before, in earlier \gerda\ \onbb\ decay analyses such as \cite{Agostini:2019hzm}, partitioning was not performed, and data from multiple detectors were combined to form a dataset for each detector type.
In the case of the \gerda\ Phase~II data, very few events (in fact, only one) are observed close to \Qbb, so using a signal model of a Gaussian with an effective resolution as in Sect.~\ref{sssec:bkgeffres} is not appropriate.
Instead, a simple weighted average of the partition resolutions at \Qbb\ gives the resolution expectation value to be associated with events in the region of interest, i.e.:

\begin{align}
  \sigma = \sum_i w_i \sigma_i,
\end{align}
where the sum goes over the partitions with resolutions $\sigma_i$ and weights $w_i$.
For the three detector types we obtain the resolutions at \Qbb\ as given in Tab.~\ref{tab:typeres}.

\begin{table}[h]
  \centering
  \begin{tabular}{c|c}
    Detector type & Resolution at \Qbb\ [keV]\\
    \hline
    BEGe & 2.8 $\pm$ 0.3 \\
    coaxial & 4.0 $\pm$ 1.3 \\
    IC & 2.9 $\pm$ 0.1 \\
  \end{tabular}
  \caption{FWHM resolutions at \Qbb\ for datasets of each detector type, reported as exposure-weighted averages.
The uncertainty is given by the standard deviation among the detector partitions.
  }
  \label{tab:typeres}
\end{table}


\section{Energy resolution uncertainty at \qbb}
\label{sec:sys_res}

The statistical uncertainty on the energy resolution decreases with rising statistics over time, and is on the order of only a few eV.
As such, the uncertainty on the energy resolution is dominated by systematic effects.
We consider various sources of systematic uncertainty, given here in decreasing order of their contribution: (i) resolution shifts over time; (ii) energy scale shifts over time; (iii) choice of the resolution fitting function.
Due to the nature of these uncertainties, their magnitude will not decrease over time, but could change if the detector setup or analysis methods change.

In the following sections, we explain how individual contributions to the systematic uncertainty were determined (Sect.~\ref{sssec:resstab} to Sect.~\ref{sec:choice_of_resolution_function}), and how they are combined together to give a total uncertainty per partition (Sect.~\ref{sec:total_resolution}).

\subsection{Resolution stability}
\label{sssec:resstab}

We consider a systematic uncertainty estimated from the fluctuations in the resolution obtained for each calibration over time.
For each partition, we calculate the standard deviation of the resolution at FEP, $\sigma_\mathrm{FEP}$, among individual calibration runs in that partition.
Assuming that in Eq.~\ref{eq:rescurve}, any systematic fluctuation of the energy resolution is caused by the two correlated parameters changing proportionally, the energy resolution uncertainty $\delta$ divided by the energy resolution $\sigma$ is independent of energy.
This is supported by the high degree of correlation between the fit parameters $a$ and $b$ of Eq.~\ref{eq:rescurve} of $-0.81$ for the fitted partition resolution curves.
With this specific model, we can translate the uncertainty at the FEP energy to \qbb:

\begin{align}
\delta_{Q_{\beta\beta}} = \frac{\sigma_{Q_{\beta\beta}}}{\sigma_\mathrm{FEP}} \delta_\mathrm{FEP}.
\end{align}
The mean value for this component across all partitions is 0.11\,keV, with a standard deviation of 0.06\,keV.

\subsection{Pulser stability}
\label{sec:pulserstab}

Once the energy scale has been determined via a calibration as described in Sect.~\ref{subsec:peak_params}, the calibration curves are used until the next calibration. While several parameters are monitored to ensure detector stability, fluctuations of the energy scale can still deteriorate the resolution for physics data compared to calibration data.  Fluctuations on time scales smaller than the typical calibration duration (1.5\,h) are also present in the calibration data.
The  effect from these short-term fluctuations will thus be included in the calculated effective resolution.
Fluctuations on larger time scales, up to around one week, can, within the restraints of our data quality requirements, contribute additionally to the resolution in physics data compared to the resolution obtained from calibration data.

This additional contribution was estimated using the position of test pulser events (see Sect.~\ref{sec:monitoring}). Shifts in the test pulser positions averaged over 1.5\,h, normalised by their statistical uncertainty, were analyzed.
Were the variation in energies due only to statistical fluctuations, these normalized residuals would be distributed normally with a mean of 0 and a standard deviation of 1.
The observed deviation from this standard normal distribution can be quantified as an additional contribution to the resolution, which is typically on the order of 0.2\,keV (1$\sigma$) or 0.6\,keV (FWHM).
As an example, for a detector partition with a resolution of \linebreak$\mathrm{FWHM}=3.0$\,keV, the additional systematic uncertainty is given by:

\begin{align}
	\delta_\mathrm{sys.} &= \sqrt{\mathrm{FWHM}^2 + \mathrm{(0.6\,keV})^2} - \mathrm{FWHM}\nonumber\\
	&= 0.06\,	\mathrm{keV}.
\label{eq:pulsersys}
\end{align}
The mean value for this component is 0.08\,keV, with a standard deviation of 0.07\,keV among partitions.

\subsection{Choice of the resolution function}
\label{sec:choice_of_resolution_function}

We used the square root of a linear function to model the resolution as a function of energy (Eq.~\ref{eq:rescurve}).
While this choice is physically well-motivated, including both statistical variations in the number of charge carriers, and effects due to the electronics, there are some common alternatives.
For example, one could add a quadratic term under the square root to model the effects of incomplete charge collection or integration,

\begin{align}
	\sigma = \sqrt{a + bE + cE^2}.
\label{eq:rescurvequad}
\end{align}
To estimate the variation of the resolution at \qbb\ for the different choices of functions, the values obtained for the two discussed choices are compared. Using the square root of linear (Eq.~\ref{eq:rescurve}) and quadratic (Eq.~\ref{eq:rescurvequad}) functions, an average difference of 0.05\,keV is obtained, with a standard deviation of 0.05\,keV among partitions.

\subsection{Total resolution uncertainty by partition}
\label{sec:total_resolution}

The total resolution uncertainty is obtained by summing individual contributions in quadrature, thereby assuming no correlations.
The resultant FWHM resolution uncertainties at \Qbb\ of the partitions vary between 0.04\,keV and 0.37\,keV, with a mean (standard deviation) of 0.13 (0.07)\,keV.

\section{Energy bias and uncertainty}
\label{sec:sysscale}

Due to the different assumptions and approximations in the calibration procedure, slight biases in the energy scale may remain.
Such biases may, for example, be caused by the integral non-linearity of the FADCs~\cite{Abgrall:2020jto}.
Small non-linearities in the energy scale are for example neglected due to the use of a linear calibration function. Therefore a peak from a $\gamma$ ray with well defined energy might be displaced towards higher or lower energies. Correspondingly, for each individual event, while its reconstructed energy will fluctuate according to the resolution, it may also be systematically displaced.

To evaluate the energy bias per partition near \qbb, we look at the residual at the SEP defined in Sect.~\ref{sec:monitoring} in the combined calibration spectrum, since the SEP is very close to \qbb\ with a difference of 64.5\,keV.
The statistics is sufficient to reach a precision of $\mathcal{O}$(0.01\,keV) for the SEP position.
The average bias is found to be $-0.07$\,keV, with a standard deviation of 0.29\,keV among the partitions. Since the \onbb\ decay search is extremely sensitive to the energy of the events close to \qbb, in the final \gerda\ analysis \cite{Agostini:2020xta}, we correct for the energy bias of the events that fall into the energy range considered for the \onbb\ decay search (1930\,keV to 2190\,keV), by adding the amount of bias to the calibrated event energy.
This approach is justified by studying the residuals at the $^{42}$K peak (1525\,keV) and the DEP (1592.5\,keV), which are two closely located peaks with the former appearing in the physics data~\cite{Abramov:2019hhh} and the latter in the calibration data. The relation between them is consistent with that between \(Q_{\beta\beta}\) and the SEP.

For the uncertainty of the bias, we use the residual fluctuations of the SEP over time near \qbb.
We additionally include a systematic uncertainty of 0.02\,keV accounting for the potential difference between the bias at SEP and that at \qbb. It was estimated by performing a linear interpolation between the residuals at the DEP and the SEP which are on the two sides of \qbb.
In total, the average bias uncertainty is 0.17\,keV.

\section{Comparison to physics data}
\label{sec:compphy}

The two strongest $\gamma$ lines in our physics data spectrum are those due to $^{40}$K (1460.8\,keV) and $^{42}$K (1524.7\,keV) decays \cite{Agostini:2020xta}. The measured resolution of these peaks allows for a cross-check to the conclusions drawn solely from calibration data. For every partition, the background energy spectrum around each of these lines is fitted using a Gaussian for the signal and a linear function for the background.
The background rate was constrained to be non-negative across the fitting window.
Partitions with potassium peaks with low counting statistics, i.e. those whose best-fit is compatible with zero counts, are excluded from further analysis.

Given their proximity in energy, the extracted resolution for each of the two lines is expected to coincide within 0.05\,keV.
Indeed, no significant difference between the resolutions of the two peaks was found.

\begin{figure}[thbp!]
\centering
\includegraphics[width=0.5\textwidth]{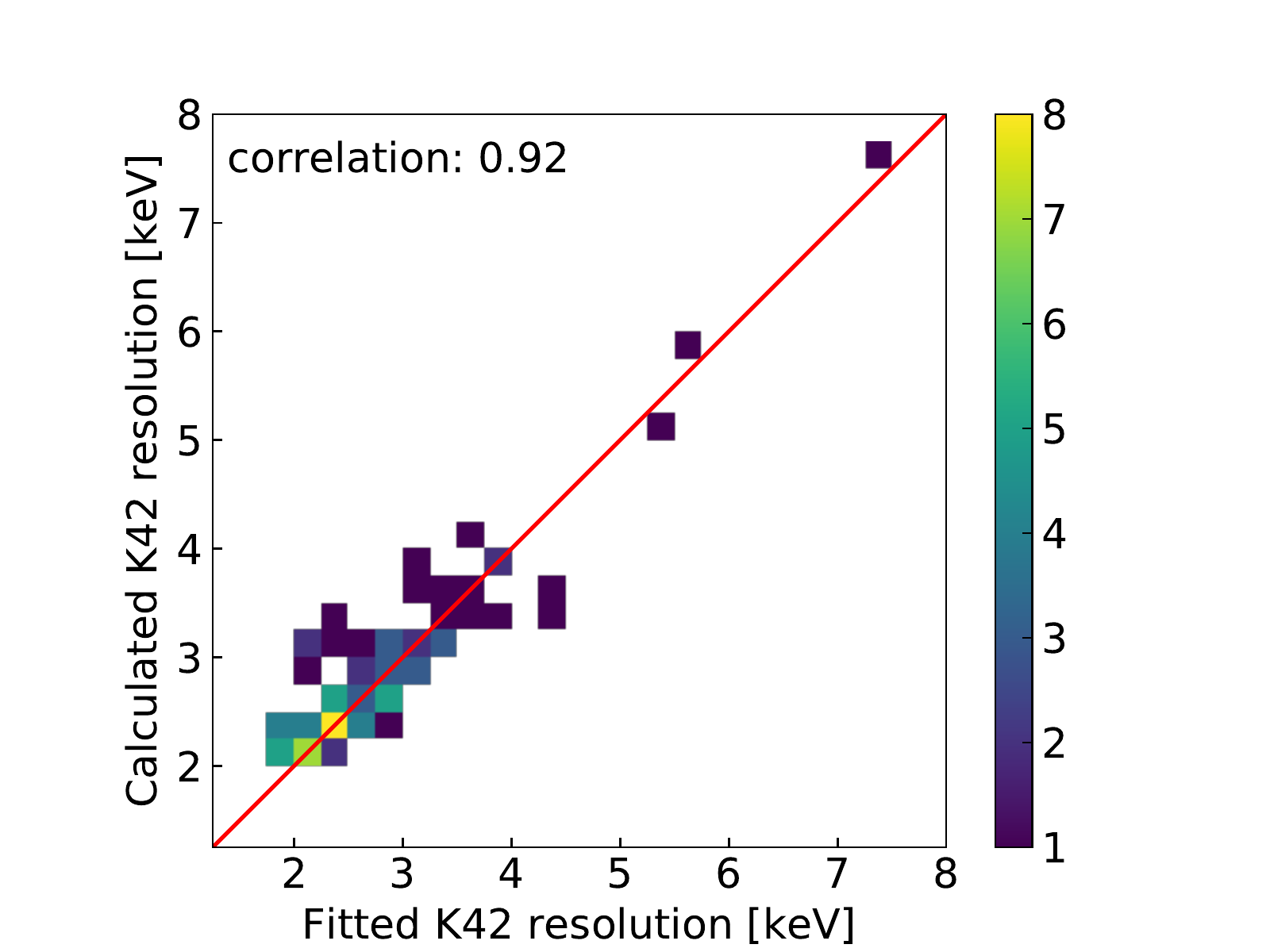}
\caption{Resolution of the 1524.7\,keV $^{42}$K line as measured from physics data and extracted from calibration data, for each detector partition.
  The red line shows the case of perfect agreement.
}
\label{fig:Kressys}
\end{figure}

We compared the resolution obtained in the potassium lines with the one predicted from the resolution curves extracted from the combined energy spectra (see Sect.~\ref{sec:combinedanalysis}), as shown in Fig.~\ref{fig:Kressys}.
The systematic uncertainty for the calibration resolution is calculated in the same way as described in Sect.~\ref{sec:sys_res}.
The measured resolutions and predicted values from calibration data show a high degree of correlation, with a Pearson correlation coefficient of 0.92, and with 66\% compatibility within one $\sigma$.
Similar results are obtained for the $^{40}$K line.

\section{Conclusions}
\label{sec:conclusions}
A reliable and stable energy scale is crucial to the search for \onbb\ decay of $^{76}$Ge performed with the \gerda\ experiment.
The event energies are reconstructed using the ZAC filter to minimize the effects of low-frequency noise.
To  preserve the excellent energy resolution of the germanium detectors when combining data over a long period of time, they are calibrated weekly using \Th\ sources. By identifying $\gamma$ peaks in the recorded spectrum the energy scale and energy resolution can be determined.

For each calibration, the stability of the energy scale and resolution is monitored via the 2.6\,MeV FEP from \Tl\ decays. Between successive calibration runs the energy scale is monitored via test pulser events injected into the readout electronics of the HPGe detectors. Data with short-term instabilities are discarded from further analysis.

To more accurately reflect the properties of a detector at a certain time, we have introduced the partitioning of the detectors' data into stable sub-periods.
The stability is based on the long-term changes of the energy resolution at the FEP and the residual at the SEP.

For each partition, a combined calibration analysis is performed to calculate the energy resolution used for the \onbb\ decay analysis.
For this purpose, calibration data in a partition are combined into a single spectrum. The resolution curve is obtained by fitting a resolution model function to the obtained resolutions of individual peaks in the combined spectrum.
Among the partitions, the calculated resolutions at \Qbb\ range from 2.3\,keV to 8.8\,keV, with an exposure-weighted mean (standard deviation) of 3.0(0.8)\,keV.

Alternatively, effective resolution curves per detector type are calculated by modeling the signal by a single Gaussian with a width according to the standard deviation of a Gaussian mixture of the individual detector partition contributions.
Over Phase~II we obtained exposure-weighted average resolutions at \Qbb\ for the BEGe/coaxial/IC detectors of ($2.8\pm0.3$)\,keV, ($4.0\pm1.3$)\,keV, and ($2.9\pm0.1$)\,keV respectively.

Dedicated studies were performed to determine the resolution systematic uncertainties for the \onbb\ decay analysis.
Various sources of systematic uncertainty on the resolution were considered: the fluctuations of the resolution and energy scale over time, and the choice of resolution function.
The average total systematic uncertainty across all partitions is 0.13\,keV.

The energy bias for the events near \qbb\ is estimated and corrected based on the residual of the SEP. Among the partitions, the average bias is -0.07\,keV with a standard deviation of 0.29\,keV. The average uncertainty of these biases is 0.17\,keV.


The energy scale, partitioning, resolutions, and energy biases discussed in this paper are essential to the final search for \onbb\ decay with \gerda\ described in \cite{Agostini:2020xta}.
The success of the \gerda\ program in reaching the world's most stringent \onbb\ decay half-life constraint given by $T_{1/2}^{0\nu}> 1.8\cdot 10^{26}$\,yr at 90\% C.L, was achieved in part due to the excellent energy resolution offered by germanium detectors and the analysis described in this work.
This is an important step towards \textsc{Legend} in developing the next generation of \onbb\ decay $^{76}$Ge experiments~\cite{Abgrall:2017syy}.

\section{Acknowledgements}

The \textsc{Gerda} experiment is supported financially by
the German Federal Ministry for Education and Research (BMBF),
the German Research Foundation (DFG),
the Italian Istituto Nazionale di Fisica Nucleare (INFN),
the Max Planck Society (MPG),
the Polish National Science Centre (NCN),
the Foundation for Polish Science (TEAM/2016-2/17),
the Russian Foundation for Basic Research,
and the Swiss National Science Foundation (SNF).
This project has received funding/support from the European Union's
\textsc{Horizon 2020} research and innovation programme under
the Marie Sklodowska-Curie grant agreements No 690575 and No 674896.
This work was supported by the Science and Technology Facilities Council
(ST/T004169/1).
J.~Huang and C.~Ransom thank the UZH for the Postdoc and Candoc Forschungskredit fellowships respectively.
The institutions acknowledge also internal financial support.

The \textsc{Gerda} collaboration thanks the directors and the staff of
the LNGS for their continuous strong support of the \textsc{Gerda} experiment.



\begin{thebibliography}{10}
\providecommand{\url}[1]{{#1}}
\providecommand{\urlprefix}{URL }
\expandafter\ifx\csname urlstyle\endcsname\relax
  \providecommand{\doi}[1]{DOI \discretionary{}{}{}#1}\else
  \providecommand{\doi}{DOI \discretionary{}{}{}\begingroup
  \urlstyle{rm}\Url}\fi

\bibitem{Mohapatra:2006gs}
R.N. Mohapatra, A.Y. Smirnov, Ann. Rev. Nucl. Part. Sci. \textbf{56}, 569
  (2006).
\newblock \doi{10.1146/annurev.nucl.56.080805.140534}

\bibitem{Mount:2010zz}
B.J. Mount, M.~Redshaw, E.G. Myers, Phys. Rev. C \textbf{81}, 032501 (2010).
\newblock \doi{10.1103/PhysRevC.81.032501}

\bibitem{Agostini:2019mwn}
M.~Agostini et~al. (GERDA Collaboration), Eur. Phys. J. C \textbf{79}, 978 (2019).
\newblock \doi{10.1140/epjc/s10052-019-7353-8}

\bibitem{Domula:2017mei}
A.~Domula, M.~Hult, Y.~Kerma{\"{\i}}dic, G.~Marissens, B.~Schwingenheuer, T.~Wester,
  K.~Zuber, Nucl. Instrum. Meth. A \textbf{891}, 106 (2018).
\newblock \doi{10.1016/j.nima.2018.02.056}

\bibitem{Agostini:2017hit}
M.~Agostini et~al. (GERDA Collaboration), Eur. Phys. J. \textbf{C78}, 388 (2018).
\newblock \doi{10.1140/epjc/s10052-018-5812-2}

\bibitem{Gunther:1997ai}
M.~Gunther, et~al. (Heidelberg-Moscow Collaboration), Phys. Rev. D \textbf{55}, 54 (1997).
\newblock \doi{10.1103/PhysRevD.55.54}

\bibitem{Morales:1998hu}
A.~Morales, Nucl. Phys. B Proc. Suppl. \textbf{77}, 335 (1999).
\newblock \doi{10.1016/S0920-5632(99)00440-5}

\bibitem{Ackermann:2012xja}
K.H. Ackermann et~al. (GERDA Collaboration), Eur. Phys. J. C \textbf{73}, 2330 (2013).
\newblock \doi{10.1140/epjc/s10052-013-2330-0}

\bibitem{Agostini:2013mzu}
M.~Agostini  et~al. (GERDA Collaboration), Phys. Rev. Lett. \textbf{111}, 122503 (2013).
\newblock \doi{10.1103/PhysRevLett.111.122503}

\bibitem{Abramov:2019hhh}
M.~Agostini et~al. (GERDA Collaboration), JHEP \textbf{03}, 139 (2020).
\newblock \doi{10.1007/JHEP03(2020)139}

\bibitem{Elliott:2002xe}
S.R. Elliott, P.~Vogel, Ann. Rev. Nucl. Part. Sci. \textbf{52}, 115 (2002).
\newblock \doi{10.1146/annurev.nucl.52.050102.090641}

\bibitem{Maneschg:2017mzu}
W.~Maneschg, in \emph{{Prospects in Neutrino Physics}} (2017)

\bibitem{Baudis:2015sba}
L.~Baudis, G.~Benato, P.~Carconi, C.M. Cattadori, P.~De~Felice, K.~Eberhardt,
  R.~Eichler, A.~Petrucci, M.~Tarka, M.~Walter, JINST \textbf{10}, P12005
  (2015).
\newblock \doi{10.1088/1748-0221/10/12/P12005}

\bibitem{tarkathesis}
M.~Tarka, {Studies of Neutron Flux Suppression from a $\gamma$-ray Source and
  The GERDA Calibration System}.
\newblock Ph.D. thesis, Universit\"{a}t Z\"{u}rich (2012).
\newblock \urlprefix\url{https://doi.org/10.5167/uzh-74790}

\bibitem{baudis2013monte}
  L.~Baudis, A.D. Ferella, F.~Froborg, M.~Tarka, Nucl. Instrum. Meth. A \textbf{729}, 557 (2013)

\bibitem{Brun:1997pa}
R.~Brun, F.~Rademakers, Nucl. Instrum. Meth. A \textbf{389}, 81 (1997).
\newblock \doi{10.1016/S0168-9002(97)00048-X}

\bibitem{Agostini:2011nf}
M.~Agostini, et~al., J. Phys. Conf. Ser. \textbf{375}, 042027 (2012).
\newblock \doi{10.1088/1742-6596/375/1/042027}

\bibitem{Agostini:2011xe}
M.~Agostini, L.~Pandola, P.~Zavarise, O.~Volynets, JINST \textbf{6}, P08013
  (2011).
\newblock \doi{10.1088/1748-0221/6/08/P08013}

\bibitem{gatti1986processing}
E.~Gatti, P.F. Manfredi, La Rivista del Nuovo Cimento (1978-1999)
  \textbf{9}, 1 (1986)

\bibitem{Agostini:2015pta}
M.~Agostini et~al. (GERDA Collaboration), Eur. Phys. J. \textbf{C75}, 255 (2015).
\newblock \doi{10.1140/epjc/s10052-015-3409-6}

\bibitem{valerio:thesis}
V.~D’Andrea, {Improvement of Performance and Background Studies in GERDA
  Phase II}.
\newblock Ph.D. thesis, Gran Sasso Science Institute (GSSI) (2017).
\newblock \urlprefix\url{http://hdl.handle.net/20.500.12571/9641}

\bibitem{lazzarothesis}
A.~Lazzaro, {Signal processing and event classification for a background free
  neutrinoless double beta decay search with the GERDA experiment}.
\newblock Ph.D. thesis, Technische Universit\"{a}t M\"{u}nchen (2019).
\newblock \urlprefix\url{mediatum.ub.tum.de/node?id=1507626}

\bibitem{coldwell2016experimental}
R.L. Coldwell, G.P. Lasche, Journal of Radioanalytical and Nuclear Chemistry
  \textbf{307}, 2509 (2016)

\bibitem{litvals}
I.N.E..E. Laboratory.
\newblock {Gamma-ray spectrum catalog of isotopes}.
\newblock
  \urlprefix\url{{http://www.radiochemistry.org/periodictable/gamma_spectra/index.html}}

\bibitem{Agostini:2020xta}
M.~Agostini et~al. (GERDA Collaboration), Phys. Rev. Lett. \textbf{125}, 252502 (2020).
\newblock \doi{10.1103/PhysRevLett.125.252502}

\bibitem{Agostini:2019hzm}
M.~Agostini et~al. (GERDA Collaboration), Science \textbf{365}, 1445 (2019).
\newblock \doi{10.1126/science.aav8613}

\bibitem{hull1996charge}
E.L. Hull, J.~Xing, D.L. Friesel, R.H. Pehl, N.W. Madden, T.W. Raudorf, L.S.
  Varnell, \emph{{Charge Collection Physics in Semiconductor Detectors}} (1996)

\bibitem{chloethesis}
C.~Ransom, {Energy calibration for the GERDA and LEGEND-200 experiments}.
\newblock Ph.D. thesis, Universit\"{a}t Z\"{u}rich (2021)

\bibitem{gaussianmixturethesis}
C.~Am{\'e}ndola, {Algebraic Statistics of Gaussian Mixtures}.
\newblock Ph.D. thesis, Technische Universit\"{a}t Berlin (2017).
\newblock \urlprefix\url{https://depositonce.tu-berlin.de/handle/11303/7284}

\bibitem{Abgrall:2020jto}
N.~Abgrall et~al. (MAJORANA Collaboration), IEEE Trans. Nucl. Sci. \textbf{68}, 359 (2021).
\newblock \doi{10.1109/TNS.2020.3043671}

\bibitem{Abgrall:2017syy}
N.~Abgrall et~al. (LEGEND Collaboration), AIP Conf. Proc. \textbf{1894}, 020027 (2017).
\newblock \doi{10.1063/1.5007652}

\end{thebibliography}
\end{document}